\documentclass[sigconf, nonacm]{acmart}
\usepackage{quoting}
\quotingsetup{font=itshape,vskip=2pt, leftmargin=8pt}

\usepackage{subcaption}  

\AtBeginDocument{%
  \providecommand\BibTeX{{%
    \normalfont B\kern-0.5em{\scshape i\kern-0.25em b}\kern-0.8em\TeX}}}

\copyrightyear{2025}
\acmYear{2025}
\setcopyright{cc}
\setcctype{by}
\acmConference[IUI '25]{30th International Conference on Intelligent User Interfaces}{March 24--27, 2025}{Cagliari, Italy}
\acmBooktitle{30th International Conference on Intelligent User Interfaces (IUI '25), March 24--27, 2025, Cagliari, Italy}
\acmDOI{10.1145/3708359.3712150}
\acmISBN{979-8-4007-1306-4/25/03}

\setcopyright{none}
\renewcommand\footnotetextcopyrightpermission[1]{}
\begin{document}

\title[PromptMap]{PromptMap: An Alternative Interaction Style for AI-Based Image Generation}

\author{Krzysztof Adamkiewicz}
\email{kadamkiewicz835@gmail.com}
\orcid{0009-0006-0730-6899}
\affiliation{%
  \institution{Lodz University of Technology}
  \city{Łódź}
  \country{Poland}
}

\author{Paweł W. Woźniak}
\email{pawel.wozniak@tuwien.ac.at}
\orcid{0000-0003-3670-1813}
\affiliation{%
  \institution{TU Wien}
  \city{Vienna}
  \country{Austria}
}

\author{Julia Dominiak}
\email{julia.dominiak@p.lodz.pl}
\orcid{0000-0001-5584-5455}
\affiliation{%
  \institution{Lodz University of Technology}
  \city{Łódź}
  \country{Poland}
}

\author{Andrzej Romanowski}
\email{andrzej.romanowski@p.lodz.pl}
\orcid{0000-0001-5241-0405}
\affiliation{%
  \institution{Lodz University of Technology}
  \city{Łódź}
  \country{Poland}
}

\author{Jakob Karolus}
\email{jakob.karolus@dfki.de}
\orcid{0000-0002-0698-7470}
\affiliation{%
  \institution{German Research Center for Artificial Intelligence}
  \city{Kaiserslautern}
  \country{Germany}
}

\author{Stanislav Frolov}
\email{stanislav.frolov@dfki.de}
\orcid{0000-0003-2700-5031}
\affiliation{%
  \institution{German Research Center for Artificial Intelligence}
  \city{Kaiserslautern}
  \country{Germany}
}

\renewcommand{\shortauthors}{Adamkiewicz, et al.}

\begin{abstract}
Recent technological advances popularized the use of image generation among the general public. Crafting effective prompts can, however, be difficult for novice users. To tackle this challenge, we developed PromptMap, a new interaction style for text-to-image AI that allows users to freely explore a vast collection of synthetic prompts through a map-like view with semantic zoom. PromptMap groups images visually by their semantic similarity, allowing users to discover relevant examples. We evaluated PromptMap in a between-subject online study ($n=60$) and a qualitative within-subject study ($n=12$). We found that PromptMap supported users in crafting prompts by providing them with examples. We also demonstrated the feasibility of using LLMs to create vast example collections. Our work contributes a new interaction style that supports users unfamiliar with prompting in achieving a satisfactory image output.
\end{abstract}



\begin{teaserfigure}
  \includegraphics[width=\textwidth]{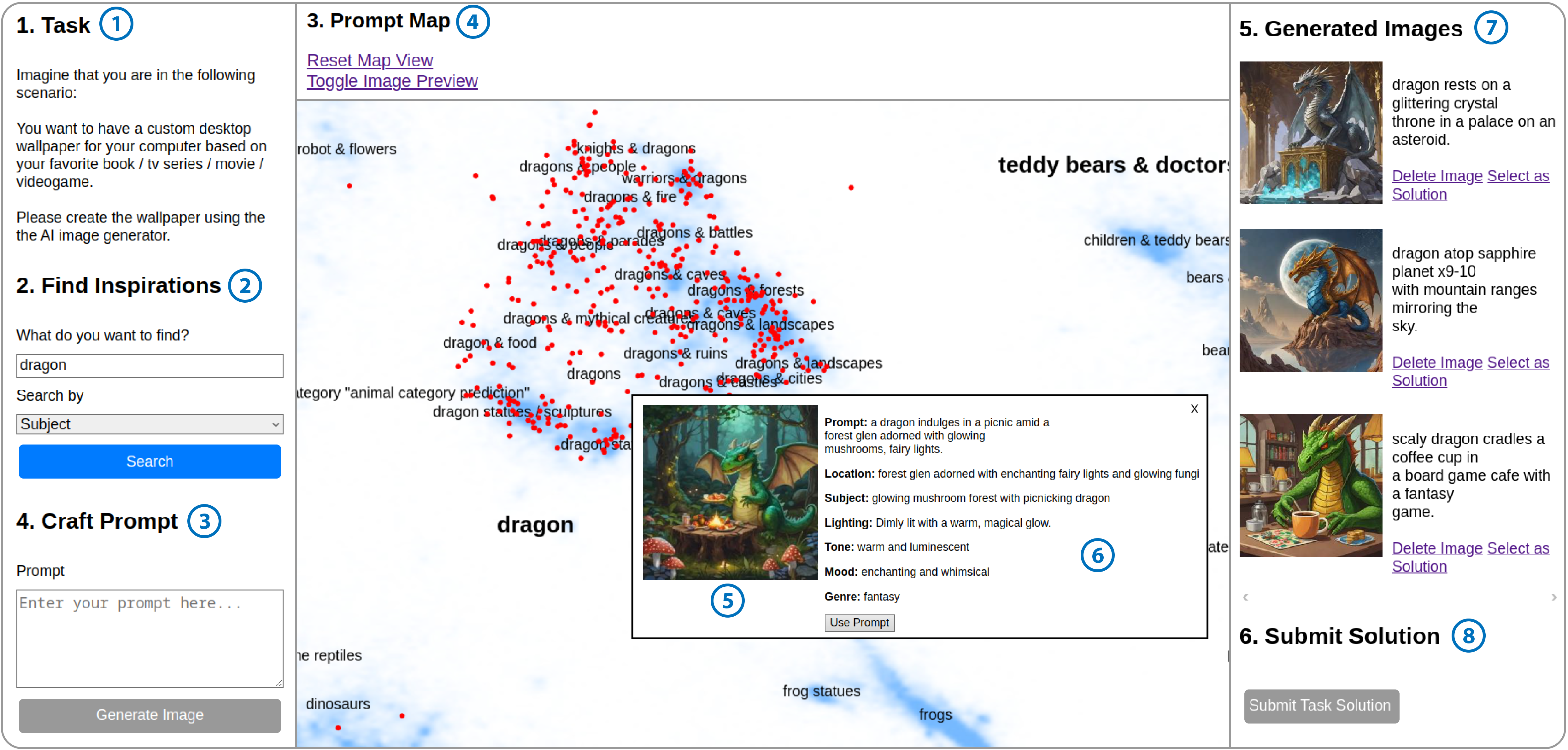}
  \caption{\emph{PromptMap} helps users craft effective prompts for generating their desired images by allowing them to explore a vast, synthetic collection of examples. The interface is composed of the following elements: (2) a search feature that visually highlights relevant examples on the map view, (3) a prompt input field for the text-to-image model, (4) a map view for exploring examples, visually grouped by topics and (5) a history of previously generated images. Each example has a preview of the output generated from the prompt (6) and predicted image attributes (7), such as the image's location, lighting, or mood. Task (8)  submission button and task description (1) were included to suit the needs of the online study.}
 \Description[User Interface]{This image illustrates an interface for creating AI-generated custom desktop wallpapers based on user preferences. It is divided into several sections: 1. Task: The leftmost panel introduces the scenario where the user is tasked to design a custom desktop wallpaper inspired by a favorite book, TV series, movie, or video game; 2. Find Inspirations: Below the task section, users can search for inspiration by typing keywords (e.g., "dragon") and selecting a search category (e.g., "Subject"). A "Search" button facilitates this process, 4. Craft Prompt: a prompt input field for the text-to-image model users can craft specific prompts in the text box provided and click "Generate Image" to initiate image creation; 3. Prompt Map: The central map visualizes various visualizations of generated prompts represented as red dots and blue clusters. Related topics cluster together, such as "dragons and forests," "dragons and cities," and "dragons and mythical creatures." Hovering over a dot displays a preview of the associated image and its prompt details; prompt Preview: A popup window in the map area provides detailed information about a selected prompt, including the description, subject, lighting, tone, mood, genre, and location; 5. Generated Images: The right panel shows the generation history. Users can delete or select an image as a solution, Submit Solution: At the bottom right, users can finalize and submit their solution after selecting a preferred image.}
  \label{fig:teaser}
 \end{teaserfigure}



\maketitle

\section{Introduction}
Text-to-image generative models can produce high-quality images from natural language descriptions. They can be used in a wide range of creative tasks such as visual art creation~\cite{ko2023Large}, news illustration~\cite{Liu2022Opal}, or industrial design~\cite{Liu20233DALL-E}. As capabilities and accessibility improved, models such as DALL-E~\cite{ramesh2021zero}, DALL-E-2~\cite{ramesh2022hierarchical} or Stable Diffusion~\cite{rombach2022high} became popular amongst the public. 

Despite the capability advancements, text prompts, at times cryptic and unintelligible to users, remain a primary input method in many popular image generation models~\cite{Liu20233DALL-E,ramesh2021zero,ramesh2022hierarchical,rombach2022high}. Crafting the desired prompts can present difficulties, especially for beginner users~\cite{ko2023Large, Pereira2023Why} as use of specialized, often esoteric (ex. "magic keywords"), language is widespread in the text-to-image model users community ~\cite{oppenlaender2023taxonomy, dehouche2023s, dallePromptBook,wang2023diffusiondb}.

Various solutions have been proposed to support users during prompt writing. Prompt engineering studies offered more formal strategies for prompting~\cite{Liu2022Design}. Automated prompt optimization methods, designed to maximize overall image aesthetic~\cite{Hao2024Optimizing, Pavlichenko2023Best}, similarity to a given example~\cite{wen2024hard}, or specified emotional expression~\cite{Wang2023RePrompt}, have been used for prompt refinement. Additionally, various prompt engineering guides~\cite{dallePromptBook}, tools~\cite{promptInterrogator}, and extensive collections of examples~\cite{wang2023diffusiondb, lexica, civitai} have been created by the AI image generation community to provide support, reference, and inspiration. 

HCI research has also offered more user-centric approaches to interacting with AI image generators. Interactive user interfaces have been developed to support users during the prompt creation process for both LLMs~\cite{strobelt2022interactive} and text-to-image models~\cite{brade2023promptify, Feng2024promptmagician, Wang2023RePrompt, wang2024promptcharm}. Those interfaces adopt various strategies such as keyword recommendation~\cite{Feng2024promptmagician}, prompt suggestion~\cite{brade2023promptify}, prompt optimization to a specific goal (ex. aesthetic~\cite{Hao2024Optimizing}), or attention-based attribution visualization~\cite{wang2024promptcharm}.

However, we found that these improvements are still lacking as existing work still focuses on supporting users on a prompt language level. Whilst existing approaches are adequate for users who want to improve their prompt writing (cf.~\cite{Liu2022Design}), they remain challenging for users wanting to find inspiration, warranting a different interaction style for AI-based image generation models. AI image galleries~\cite{lexica, civitai} offer a different approach to supporting text-to-image model users. They provide users with a wide array of example prompts accessible through text search. Yet, they are still limited to a goal-oriented exploration strategy as image examples are not connected or clustered, forcing users to rely on the search feature to explore the example space.

To allow for targeted exploration toward a desired output image, we present PromptMap, a user interface that allows users to find inspiration by browsing a large-scale synthetic database of examples. We organize our prompt collection into a map-like view with a semantic zoom, allowing users to explore it freely. Furthermore, we devised a method for generating large-scale ($n=12.3M$) synthetic prompt collections, empowering PromptMap with a vast example space.

To evaluate PromptMap, we conducted a between-subject quantitative study ($n=60$) and a within-subject qualitative study ($n=12$). During the study, users solved practical, open-ended tasks that involved the creation of an idea and subsequent generation of an image using a text-to-image model. We compared our interface to the baseline with no examples and the nearest neighbor search of the DiffusionDB~\cite{wang2023diffusiondb} dataset. We found that the presence of examples displayed by PromptMap shifted participants' workflow from a trial-and-error approach to a more example-driven one. Participants reported that PromptMap's visual clustering of thematically related images made it easier for them to explore related ideas, and displayed examples showed greater variety than the baselines. Finally, we establish synthetic data generation as a feasible alternative to scraping when preparing vast collections of prompt examples.

The main contributions of this paper are as follows: 1) Implementation of PromptMap, a system that allows users to find inspirations in an extensive collection of prompt examples. 2) An empirical study that evaluates how user's workflow differs in the PromptMap interface. 3) A process for generating large-scale synthetic datasets of prompt examples with large language models.

Additionally, we publish our large-scale synthetic prompt dataset and provide the source code~\footnote{https://github.com/Bill2462/prompt-map} for our prototype and data generation pipeline.

\section{Related work}
This section provides an overview of relevant work related to 1) text-to-image generative models, 2) prompting methods for pretrained generative models, 3) user interfaces for text-to-image generation, and 4) exploration of design spaces.

\subsection{Text-to-Image Generative Models}
Text-to-image synthesis~\cite{zhang2023text,cao2024controllable,frolov2021adversarial} has received significant attention for its ability to generate images from natural language descriptions, opening up new possibilities for practical application.

Early models, such as generative adversarial networks (GANs) \cite{goodfellow2020generative,reed2016generative}, showcased the potential of generating plausible images but struggled with synthesizing complex scenes and often lacked fine detail and resolution.

In response, more sophisticated architectures and methods were developed to enhance the image quality \cite{karras2019style,brock2018large}. The introduction of transformer-based models marked a significant milestone in the evolution of text-to-image synthesis. Early examples like VQGAN~\cite{esser2021taming} and DALL-E~\cite{ramesh2021zero} gained widespread attention for generating high-quality images directly from text inputs.

A more recent breakthrough has been the development of latent diffusion models \cite{ho2020denoising, rombach2022high}, with models like Stable Diffusion XL \cite{podell2023sdxl} or DALL-E-2~\cite{ramesh2022hierarchical} gaining widespread attention. These models excel at producing high-resolution images from text prompts and have democratized access to state-of-the-art image generation, benefiting both researchers and artists alike.

PromptMap uses a distilled version~\cite{sauer2023adversarial} of Stable Diffusion XL~\cite{podell2023sdxl} to generate example images shown in our collection. We chose the Stable Diffusion XL turbo because it demonstrates state-of-the-art performance on image generation benchmarks and, thanks to few-step inference, has significantly lower computational cost than the base model.

\subsection{Prompting Pretrained Generative Models}

With the rise of generative AI, prompting pretrained models has become a fertile research topic.

The process of crafting prompts that accomplish a given goal became known as prompt engineering. Prompts can be generally divided into "hard" and "soft". Hard prompts are constrained to a space of valid tokens, while soft prompts operate directly in the token embedding space. Prompt engineering can be done by hand or using automated techniques such as gradient descent ~\cite{liu2023pretrain}.

Previous studies have established that manually crafting effective prompts for both LLMs and text-to-image models ~\cite{Pereira2023Why, ko2023Large, jiang2020how} is challenging. Lack of explainability~\cite{wang2024promptcharm} and unpredictability of the model behavior~\cite{ko2023Large} are significant barriers to effective interaction.

Initial insights into the behavior of text-to-image pipelines were formed by Liu et al., who conducted multiple experiments investigating the capabilities, failure modes, and human preferences in the early VQGAN + CLIP text-to-image pipeline~\cite{Liu2022Design}. Further studies~\cite{oppenlaender2023taxonomy} of the subject revealed that keywords describing the style of the image (modifiers) play a vital role in the user prompts.

To improve the degree of control over the model's output, techniques such as layout conditioning~\cite{couairon2023zero} or semantic and edge maps~\cite{zhang2023adding} have been proposed as additional control inputs to image generators. Text inversion~\cite{gal2022image}, a technique for defining custom concepts as token embeddings and fine-tuning using LoRa~\cite{gal2022image}, also aim to improve the control over qualities such as style or subject appearance.

Techniques for prompt optimization and automated generation have also been proposed. For example, Promptify~\cite{Hao2024Optimizing} rewrites the prompts to maximize the aesthetic qualities of the image while preserving the topic. Methods for finding prompts reproducing a specific image~\cite{wen2024hard} or optimizing for selected quality~\cite{cao2023beautifulprompt}, such as emotional expression~\cite{Wang2023RePrompt} of the output image, have also been developed.

Finally, large-scale prompt galleries have been created to provide examples of prompts and their outputs to the users. DiffusionDB\cite{wang2023diffusiondb} dataset collects $1.1$M prompts and $11.4$M images scraped from the stable diffusion discord channel; however, it requires software engineering expertise to use as it lacks a user interface. More accessible galleries, such as lexica.art~\cite{lexica} and civitai.com~\cite{civitai}, provide access to their collections using a user-friendly website.

PromptMap builds on the idea of AI image galleries by allowing more free exploration of the example space. We also introduce a pipeline for the synthetic generation of large-scale (>$10$M) example collections with open-source LLMs.

\subsection{UI's for Text-to-Image Generative Models}

As text-to-image generative modeling has entered the mainstream, designing user interfaces~\cite{wang2024promptcharm, brade2023promptify, Feng2024promptmagician, lexica, easyDiffusion, webui} that allow users to harness those new capabilities has become an increasingly popular topic in HCI.

Different ways of supporting the user in prompting generative models have been proposed. For instance, PromptCharm~\cite{wang2024promptcharm} uses an automatic prompt refinement technique~\cite{Hao2024Optimizing} to improve image aesthetics. It also allows users to select style keywords from a large style database and provides explanations of how different words in the prompt are attributed to the output image using heatmaps. Promptify~\cite{brade2023promptify} uses a prompt suggestion engine powered by a large language model and scatter plot-based organization of the produced images to help users iterate over their prompt more effectively. LLM-powered prompt and keyword suggestion is also adopted by Opal~\cite{Liu2022Opal}, a system for conducting news illustration with text-to-image generation. The system uses suggestions from GPT-3 and similarity search to recommend keywords describing the image's subject, style, and mood. Multiple input prompting modalities were also explored to improve the fidelity of control over the output. For example, WorldSmith~\cite{dang2023worldsmith} accepts segmentation masks and sketches as additional inputs and allows for the tiling of smaller outputs to form larger compositions. Community-maintained tools, such as Easy Difusion~\cite{easyDiffusion} or Stable Difusion WebUI~\cite{webui}, are also popular among practitioners.

While many ways to support text-to-image prompting have been proposed, interfaces for finding inspiration in vast prompt collections remain relatively unexplored.

\subsection{Exploration of Design Spaces}
PromptMap implements a map-based visual exploration of a large synthetic dataset, building on previous work describing exploration of design spaces.

Opal~\cite{Liu2022Opal}, a tool for news illustration, allows users to pick keywords from a set of recommendations, with the LLM serving as the initial filter for narrowing the search space by providing recommendations based on the news article. Luminate~\cite{suh2024luminate} generates a set of dimensions that describe the output completing a prompt, allowing users to arrange samples visually along those dimensions. Luminate also uses semantic zoom to enable easier navigation of LLM outputs.

exploration of large collections of examples was seen in works such as IdeateRelate~\cite{xu2021ideaterelate}, RecipeScape~\cite{chang2018recipescape}, DreamLens~\cite{matejka2018dreamlens} or GenQuery~\cite{son2024genquery}. IdeateRelate allows users to see the conceptual distance between their current idea and related examples from a database, supporting their idea refinement process. RecipeScape~\cite{chang2018recipescape} supports the analysis of cooking recipes at scale by providing visualizations in the form of the scatter plot with cluster boundaries and the graph of important cooking steps in the selected cluster. DreamLens~\cite{matejka2018dreamlens} allows designers to explore a large ($n=1242$) collection of generative designs to find the output that best suits their needs. The interface visualizes important design properties such as surface area or strength in addition to the interactive stacked view showing multiple 3D models at once. Finally, GenQuery~\cite{son2024genquery} is focused on using examples to support the design process as it implements several example-driven tools. The interface is centered around a search of LAION-5B~\cite{schuhmann2022laion5bopenlargescaledataset} dataset using text and image queries. Text search is supported using LLM-based query concretization, which recommends keywords based on the general text query. GenQuery allows the user to select returned images as further search queries, edit returned images by mixing them with other search results, and modify the images by adding keywords describing, for example, the style. Finally, tools such as IN-SPIRE~\cite{wise1999ecological} or BERTopic~\cite{grootendorst2022bertopic} allow for the visualization of extensive document collections on scatterplots, making use of a geographic map metaphor.

In this work, we combine the previously established techniques in a novel way. We use a custom data generation pipeline to create a vast example space, which we then present as an easily explorable, 2-dimensional map with semantic zoom.

\section{Designing a new interaction style for text-to-image models}

PromptMap offers users an alternative way of accessing a prompt example database. We introduce a new interaction style for text-to-image models that relies on an exploration of a vast space of examples organized using the geographic map metaphor.

\subsection{Creating Vast Example Space Through a Synthetic Prompt Dataset}

The main goal of the PrompMap system is to support users by providing them access to an extensive collection of relevant examples. Existing works, such as promptify~\cite{brade2023promptify}, use human prompts scraped from online sources. For instance, DiffusionDB~\cite{wang2023diffusiondb} dataset was obtained by scraping the official stable diffusion discord channel.

However, using human-generated prompts comes with several limitations. The presence of NSFW content is a known problem of human prompts ~\cite{chen2023twigma}. Additionally, prompts from online galleries frequently contain a large number of various keywords (ex., '4k', 'artstation', 'greg rutkowski'). Reliance on such keywords can lead to prioritization of surface-level aesthetical qualities and reduce the overall diversity and innovation ~\cite{palmini2024patternscreativityuserinput}. Furthermore, collecting prompts in the wild can be challenging due to the heavy use of fine-tuned models in the community. Fine-tuning makes it easier to achieve certain effects but can also introduce new vocabulary or remove the need to specify characteristics, such as style or main subject \cite{ruiz2023dreamboothfinetuningtexttoimage}. Finally, there is evidence that the prompting style may change with model capabilities \cite{jahani2024generativemodelsimproveadapt}. Scraped prompts, therefore, have to be filtered by the used model, which cuts back on the number of available samples.

In this work, we address those issues by generating a large-scale ($n=12.3$M) synthetic prompt dataset (\autoref{sec:syn_data}). Our dataset ($538$ per $10$k) shows an $8$ times lower number of NSFW detections compared to human prompts in DiffusionDB ($4357$ per $10$k) in addition to a higher number of unique image subjects. Our custom prompt generation process also allows for easy customization of the prompt style by simply modifying LLM prompts that form the part of the pipeline.

We believe that access to a diverse and high-quality collection of examples will help the user achieve their creative goals. As such, we also confirmed that our dataset contains a wide variety of unique examples.

\begin{figure*}
  \includegraphics[width=\textwidth]{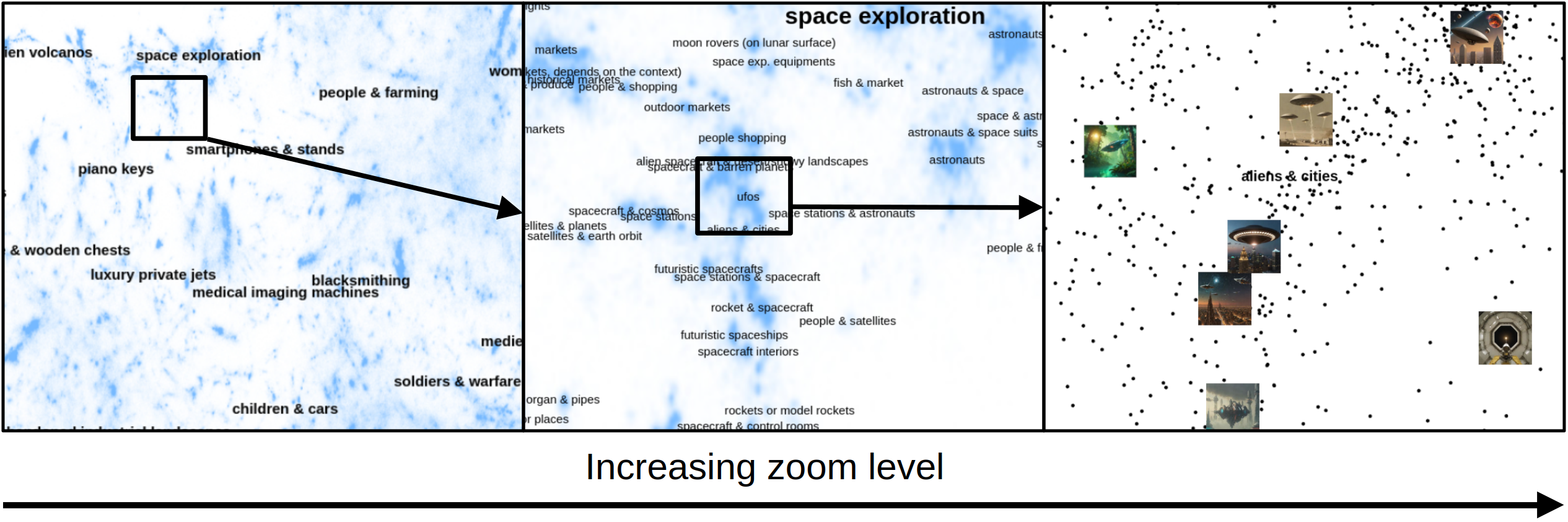}
  \caption{Map view shown in the \emph{PromptMap} interface implements semantic zoom. As the user zooms in, more labels appear. At the highest levels of zoom, the density map fades out and is replaced by examples represented as individual points. The blue color indicates density. Samples with similar topics form clusters and are visible as darker blobs on the map.}
  
 \Description[Semantic zoom]{Image demonstrates an interactive zooming process on a visual map of AI-generated prompts. It highlights how increasing zoom levels allow users to explore specific topics more deeply. The Figure is divided into three panels: Initial Zoom Level (Left): The first panel shows a broad view of the map, with general categories like "space exploration,' 'piano keys,' and 'luxury private jets.' Topics are distributed sparsely, and users can identify larger clusters representing thematic areas of interest. Intermediate Zoom Level (Middle): The second panel provides a closer view of the "space exploration" cluster. Subtopics like "satellites and earth orbit," "aliens and barren planets," and "futuristic spacecraft" become visible. The granularity increases, and users can explore more specific subcategories within the more prominent theme. Highest Zoom Level (Right): The final panel zooms further into a specific subcategory, such as 'aliens and cities.' Individual prompts, represented as dots, are visible, along with thumbnail previews of corresponding AI-generated images. These images provide detailed visual representations related to the chosen subtopic. At the bottom, an arrow with a label indicates the progression, reading 'Increasing zoom level', signifying the transition from broad categories to specific prompts and visual outputs. This visualization showcases the detailed exploration capabilities of the interactive map, enabling users to drill down from general ideas to particular creative inspirations.}
  \label{fig:semantic_zoom}
\end{figure*}

\subsection{Enabling Structured Exploration Through a Map View}

Existing platforms that allow for exploration of examples, such as lexica.art~\cite{lexica} or civita.ai~\cite{civitai}, display images as an infinitely scrollable grid. It is also worth noting that both platforms heavily rely on search for exploration and do not contain detailed indexes of topics.

Contrarily, PromptMap encourages exploration by organizing the examples into a spatial map representation with semantic zoom (\autoref{sec:map_view}). We draw inspiration from how geographic map metaphor can visualize document topics and allow for easy identification of similar samples using their spatial proximity ~\cite{wise1999ecological,grootendorst2022bertopic}. The examples on our map are represented by points placed according to the main topic of the image. PromptMap uses a density map, labels, and a search engine to help users navigate the map and locate areas of interest.

\section{Implementation}

In this section, we introduce the implementation of PromptMap. Specifically, we discuss 1) Map View, 2) Search, 3) Backend, and 4) Data Generation Procedure.

\subsection{Map View}
\label{sec:map_view}

To allow users to explore our synthetic prompt collection more freely, we organize it as a two-dimensional map view. \autoref{fig:teaser} shows the user interface and its main components. Each prompt example is represented as a black point. Point positions are determined based on the main subject of the image. Samples with the same or similar main subjects appear close to each other, while samples with dissimilar main subjects are placed far apart. This results in topics forming distinct, high-density regions, aiding in the exploration. Users can navigate the map by dragging it with their mouse and changing the zoom level with their mouse scroll wheel. The background color indicates where the points are located. The darker the color, the higher the density of examples. Clusters of samples with similar topics are visible as darker blobs. Text labels indicate the approximate subject of samples In the given area. Figure \ref{fig:semantic_zoom} shows the operation of semantic zoom. As the user zooms in, more labels will appear, and the density map eventually fades out and is replaced by the view with points. a For randomly selected fraction points, the image previews is displayed as icons.

Users can hover their mouse over the points and see the prompt, an example image generated from the prompt, and prompt annotations describing the likely location, subject, lighting, tone, mood, and genre of the output image. Prompt annotations used in our examples are inspired by prompt keyword classification~\cite{dehouche2023s} used by text-to-image model users. When a user clicks on the point, the window stays open, allowing them to copy the prompt into the clipboard.

\begin{figure*}[h]
  \includegraphics[width=\textwidth]{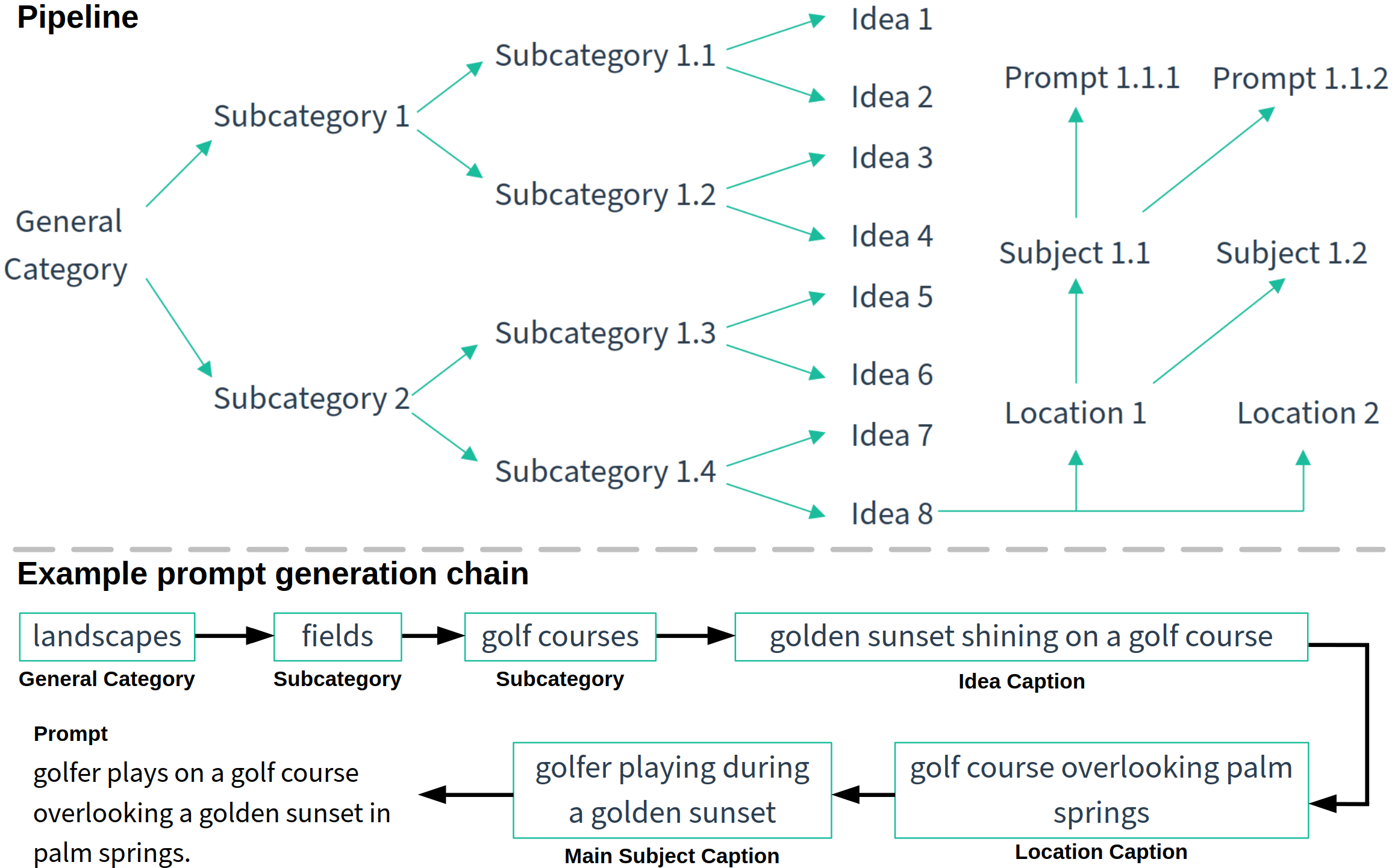}
  \caption{\emph{PromptMap} relies on the recursive expansion of concepts to generate a large dataset of prompts with an LLM. We initialize the generation with $160$ general categories describing images obtained from the GPT4-o model. The first three expansion stages create captions describing the ideas for images. We ask the LLM to find subcategories of general categories and then subcategories of those subcategories. Then, for each subcategory, we generate several ideas. We continue recursive expansion by prompting the LLM for $10$ location captions per idea, and then for each location and parent idea; we prompt for $5$ main subject captions. Finally, we prompt LLM to merge the subject and its parent location and idea into a prompt. This process generates $12.3$M prompts from the initial set of categories. For clarity, deduplication was omitted from this plot, and only two outputs are shown for each input.}
 \Description[Diagram of Generation Pipeline]{This image illustrates a two-part conceptual framework for generating prompts using a hierarchical structure, divided into the following sections:
 Pipeline (Top Section): The diagram starts with a General Category, which branches into various Subcategories (e.g., Subcategory 1 and Subcategory 2). Each subcategory further splits into smaller Subcategories (e.g., Subcategory 1.1, Subcategory 1.2). These smaller subcategories lead to Ideas (e.g., Idea 1, Idea 2), concrete inspirations, or themes derived from the initial category. One example idea is broken down into specific Locations, next to Subjects, and  Prompts. This hierarchy represents how a general concept evolves into detailed creative prompts.
 Example Prompt Generation Chain (Bottom Section):
 This section provides a practical example of the pipeline: Starting with a General Category ("landscapes"), the hierarchy narrows down to a Subcategory ("fields") and then further into another Subcategory ("golf courses"). The chain ends with an Idea Caption ("golden sunset shining on a golf course"). The prompt derived from this chain is: "A golfer plays on a golf course overlooking a golden sunset in Palm Springs." The Main Subject Caption and Location Caption refine the description further, ensuring specificity (e.g., "golfer playing during a golden sunset" and "golf course overlooking Palm Springs"). This Figure visualizes the creative process of generating structured, detailed prompts from broad concepts.}
  \label{fig:data_gen_scheme}
\end{figure*}

Inspired by related works such as Promptify~\cite{brade2023promptify} or PromptMagician~\cite{Feng2024promptmagician}, we determine point positions using dimensionality reduction of embeddings. We use a state-of-the-art Vision-Large-Language Model~\cite{yao2024minicpm} (VLLM) to annotate the main subject of each example image. We then compute the text embeddings of those annotations using all-mpnet-base-v2 text embedding model~\cite{reimers-2019-sentence-bert} and perform dimensional reduction using UMAP~\cite{2018arXivUMAP, mcinnes2018umap-software} algorithm. The color background indicating point density is a $2000x2000$ 2D histogram obtained by binning the point positions.

An expert familiar with the outputs from the UMAP algorithm manually selected the label positions. We then picked the $20$ nearest neighboring samples to each position and prompted LLM to generate the label that best fits the map contents at a given location based on the main subject captions.

\subsection{Search}

To aid in exploring the prompt collection, we implemented a nearest-neighbor search feature to find samples by prompt annotations. For example, users can search for 'lush forest' by location annotation and see examples of prompts that produce scenes in a lush forest. Search results are displayed as red points as seen in \autoref{fig:teaser}. The search result points behave in the same way as regular points with the exception of being visible on all levels of the zoom.

The search feature uses the Inverted File Product Quantization (IVFPQ) algorithm implemented by the faiss~\cite{faiss} software package. We compute the text embedding of each prompt annotation caption using an all-mpnet-base-v2 text embedding model and generate the IVFPQ index separately for each annotation type. Search queries from the user are encoded to embedding by the text embedding model, and then $200$ nearest neighbors are returned from the search index corresponding to the selected 'search by' setting. Due to the use of an efficient IVFPQ approximate nearest neighbors algorithm, the search runs in real time despite the large dataset size.

\subsection{Creating Images With PromptMap, User Scenario}

Anna, who works in an office, decides to host a BBQ for her friends to celebrate the start of summer. She wants to create a special invitation to send out to everyone but has no idea how to design something nice. Anna has heard of AI image generation tools, but whenever she tried them before, she felt lost and frustrated. Writing good prompts eluded her, and she always struggled to achieve a good result.

She decides to try out PromptMap to generate the BBQ invitation. Instead of formulating a long, detailed prompt, Anna starts with a simple idea. She types in BBQ in PromptMap's search and immediately sees a map filled with different images. She begins browsing through the clusters and quickly comes across pictures of garden parties, tables of food, and people enjoying time outdoors.

Anna decides to take it a step further. She searches again, this time for a "flower-filled garden" because she wants her invitation to have a pleasant, summery vibe. She sees several images of greenery and flowers in the background and knows it is perfect. She chooses one of the images, takes the prompt information provided by PromptMap, and adds parameters for a BBQ and smiling people, which she had come across beforehand. After some final tweaks, like switching the style to something more playful and cartoonish to give the invitation a light and funny feeling, she is satisfied with the results. She starts printing the image on her invitation booklets.

\subsection{Backend Implementation}

The UI is implemented in the form of a web application. The backend is written in python3 language and uses Flask for serving the required APIs. The front end is written in JavaScript. We use Lightning Memory-Mapped Database to store and retrieve all image examples. The application runs inside a docker container on a Linux machine. We use the A10 GPU to run the stable diffusion turbo model. The rest of the system runs on the CPU without GPU acceleration. The system requires approximately 20GB of RAM and 600GB of storage for the required data.

\subsection{Synthetic Data Generation}
\label{sec:syn_data}

\begin{figure}
\includegraphics[width=\columnwidth]{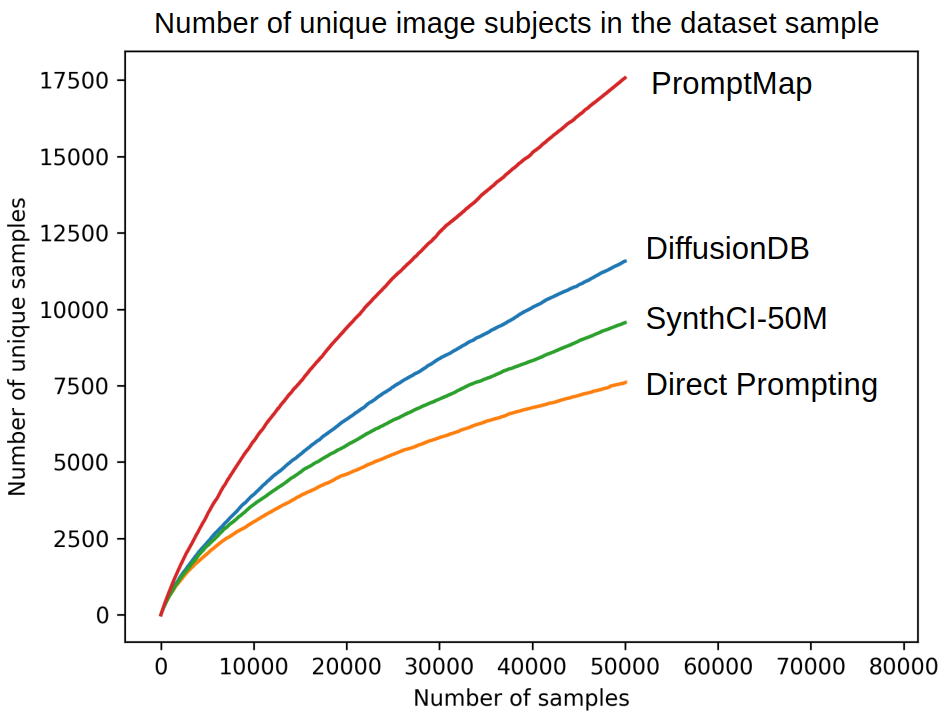}
\caption{We plot the number of unique subjects obtained through annotation of images against the number of examples in a random $50$k image sample. We compare our dataset against DiffusionDB~\cite{wang2023diffusiondb}, SynthCI-50M~\cite{hammoud2024synthclip}, and directly prompting the model to generate image captions. Our dataset demonstrates a significantly larger number of unique subjects than tested baselines.}

\Description[lineplot]{Chart compares the number of unique image subjects generated in a dataset sample across four methods as the number of samples increases. The graph provides insight into the diversity of image outputs produced by each method.
Title: "Number of unique image subjects in the dataset sample"
X-axis: Represents the number of samples, ranging from 0 to 80,000.
Y-axis: Represents the number of unique samples, ranging from 0 to 17,500.
Methods Compared: Four VLLM are represented with separate curves:
PromptMap (red curve): Achieves the highest diversity, showing the steepest growth in the number of unique subjects as sample size increases, reaching over 17,500 unique samples.
DiffusionDB (blue curve): Demonstrates moderate growth, ranking second in diversity and trailing PromptMap.
SynthCI-50M (green curve): Exhibits slightly slower growth compared to DiffusionDB, ranking third in diversity.
Direct prompting (orange curve): Shows the slowest growth and the lowest diversity, remaining below 7,500 unique samples at the maximum sample size.}
\label{fig:n_unique_samples}
\end{figure}

Unlike existing solutions~\cite{brade2023promptify, Feng2024promptmagician, wang2024promptcharm} that make use of human-written prompt collections as a part of their implementation, we generate a custom prompt dataset containing $11.3$M examples by employing open-source Mistral-7B-Instruct V0.2~\cite{jiang2023mistral} Large Language Model (LLM).
During our initial experiments, we discovered that when directly prompted to produce prompts, the Mistral LLM has a high repetition rate of image subjects, as evident from the \autoref{fig:n_unique_samples}. To combat this problem, we designed a multistage pipeline for prompt generation. Our pipeline relies on the recursive expansion of general categories to produce the final dataset. We initialize the generation by providing $160$ general categories generated using GPT4-o~\cite{achiam2023gpt} LLM. We then generate ideas for images by prompting LLM to find $10$ subcategories of initial categories and then $10$ subcategories of each generated subcategory. Finally, we prompt the LLM to generate $20$ image idea captions for each sub-sub category. Each expansion stage, in theory, multiplies the number of samples by the number of requested outputs.

In practice, different inputs can result in similar results, so to ensure the high diversity of the dataset, we employ a deduplication process. We implement it by computing CLIP text embeddings of each stage output and finding duplicates using the approximate nearest neighbors algorithm (IVFPQ). We find $200$ nearest neighbors for each sample and remove neighbors with cosine similarity larger than $0.7$.

Because of the recursive nature of the pipeline, the number of output prompts is highly dependent on the number of outputs from the first stages of the pipeline. To ensure enough subcategories are present, we combined outputs from $400$ passes over the general categories before deduplicating, which resulted in $11.6$k unique subcategories. The first three stages of the pipeline expand the $160$ general categories to $247$k ideas caption.

As visible in the example chain shown in \autoref{fig:data_gen_scheme}, the idea caption represents the general idea behind the image. The last two stages of the pipeline are then intended to add more details describing the location and main subject. This process also creates multiple variations of each idea, further increasing the sample count. For each idea, we prompt for $10$ different locations where the scene can take place. Then, for each location, we prompt for $5$ different main subjects of the scene that match the location and parent idea. Then, for each main subject, we prompt the LLM to write a prompt that matches the parent idea, parent location, and the main subject, resulting in $12.4$M prompts.

In the final step, we generate prompt annotations by asking the LLM to predict the likely lighting, mood, tone, and genre of the output image given the prompt.

To provide image examples, we used the distilled version~\cite{sauer2023adversarial} of stable-diffusion-xl-base-1.0~\cite{podell2023sdxl} to generate an example output image for each prompt. Although the rule preventing the generation of NSFW prompts was placed in the LLM prompts, we additionally ran NSFW detector~\footnote{https://huggingface.co/CompVis/stable-diffusion-safety-checker} to further reduce the chance of NSFW images being present. Samples flagged as NSFW were removed from the dataset. We found that our dataset ($4357$ per $100$k) contains significantly fewer samples flagged as NSFW compared to the DiffusionDB~\cite{wang2023diffusiondb} dataset ($538$ per $100$k) containing human written prompts.

\begin{figure}
\includegraphics[width=\columnwidth]{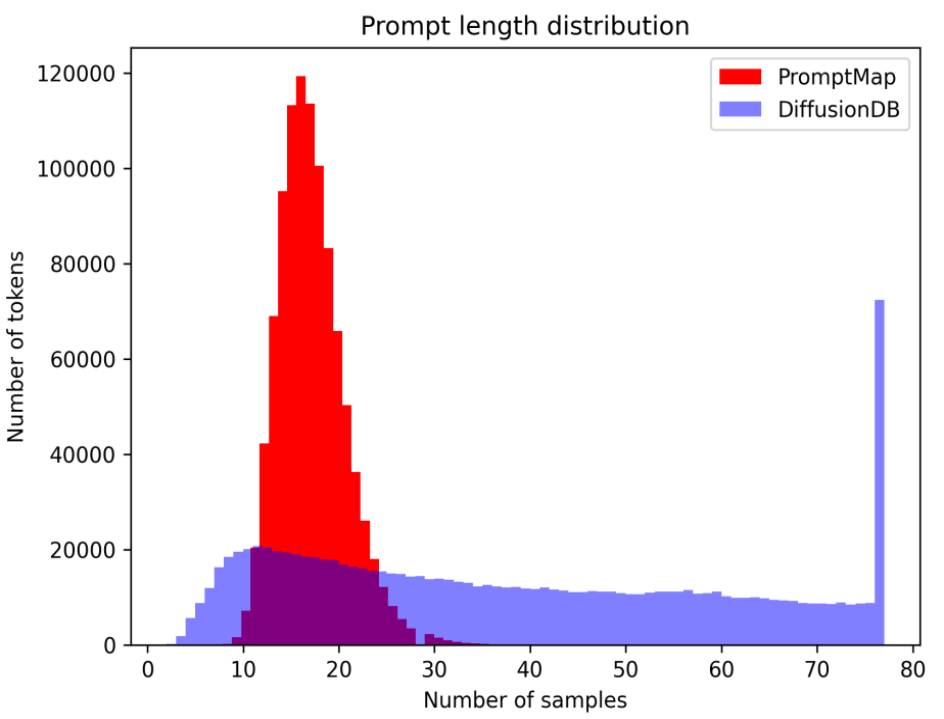}
\caption{We compare the distribution of prompt lengths between our synthetic dataset and human written prompts from DiffusionDB~\cite{wang2023diffusiondb}. We find that our prompts (M=$17.2$, SD=$3.7$) are, on average, more concise and more consistent in length than prompts in DiffusionDB (M=$38.0$, SD=$22.3$).}
\Description[Histogram]{Chart visualizes the distribution of prompt lengths in terms of the number of tokens for two methods: PromptMap and DiffusionDB.
Title: "Prompt length distribution"
X-axis: Represents the number of tokens in a prompt, ranging from 0 to 80.
Y-axis: Represents the number of samples (prompts) for each token count, reaching up to 120,000.
Legend:
PromptMap (red): Represents the token distribution for prompts generated using PromptMap.
DiffusionDB (blue): Represents the token distribution for prompts generated using DiffusionDB.
Observations:
PromptMap (Red Distribution):
The token distribution is concentrated between 10 and 20 tokens.
It peaks sharply at approximately 18 tokens, indicating a consistent and concise prompt length for most samples.
DiffusionDB (Blue Distribution):
The distribution is wider and spans across a broader range of token lengths.
While there is some overlap with PromptMap in the 10–20 token range, DiffusionDB extends beyond 30 tokens, indicating greater variability and a notable tail in the distribution.}
\label{fig:prompt_len}
\end{figure}

We benchmark our dataset by counting the number of unique subjects as a function of the number of samples in the test sample. We draw $50$k random prompts from the PromptMap dataset and state-of-the-art in the area of prompt datasets: DiffusionDB~\cite{wang2023diffusiondb} and SynthCI-50M~\cite{hammoud2024synthclip}. Additionally, we generate 50k prompts by directly asking the Mistral-7B-Instruct V0.2 LLM to create diverse image captions. 

We generate a single image per prompt with the stable-diffusion-xl-turbo model and use VLLM~\cite{yao2024minicpm} to annotate the main subject of each image. We then compute the text embeddings with the all-mpnet-base-v2 text encoder and follow the deduplication procedure used in the pipeline. Finally, we count the number of unique samples as a function of the sample count.

The results, visible in Figure \autoref{fig:n_unique_samples}, show that the pipeline adopted in PromptMap demonstrates the highest generation rate of image examples with unique subjects. On the other hand, the strategy of directly prompting the LLM demonstrates the lowest rate of unique subject generation. SynthCI-50M~\cite{hammoud2024synthclip} dataset, which used concept bank obtained from WordNet Synsets and Wikipedia common unigrams, bigrams, and titles, performed worse than human prompts but better than the direct prompting for image captions. Prompts sampled from DiffusionDB~\cite{wang2023diffusiondb} perform better than the SynthCI dataset but remain worse than our dataset.

Additionally, we compare the distribution of prompt lengths (in token counts) between the DiffusionDB and PromptMap. \autoref{fig:prompt_len} shows that, on average, prompts in PromptMap (M=$17.2$, SD=$3.7$) are more concise than prompts in DiffusionDB (M=$38.0$, SD=$22.3$). Prompts in PromptMap also demonstrate significantly higher length consistency than human-written prompts in DiffusionDB.

\begin{figure*}[ht]
    \begin{subfigure}[t]{0.30\linewidth}
        \includegraphics[width=\linewidth]{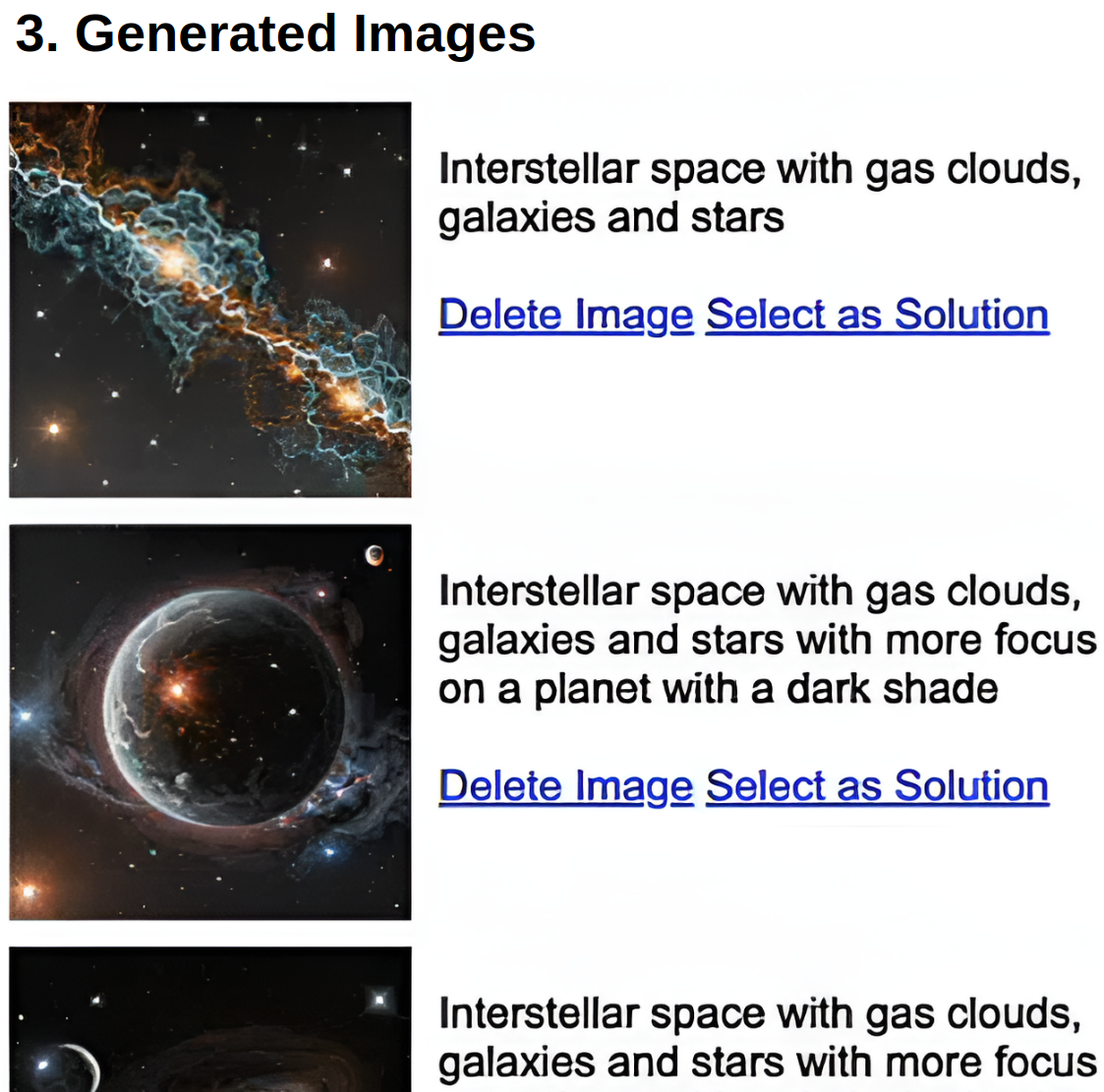}
        \caption{\textsc{No Support}}
        \label{fig:no_support}
    \end{subfigure}
    \hfill
    \begin{subfigure}[t]{0.30\linewidth}
        \includegraphics[width=\linewidth]{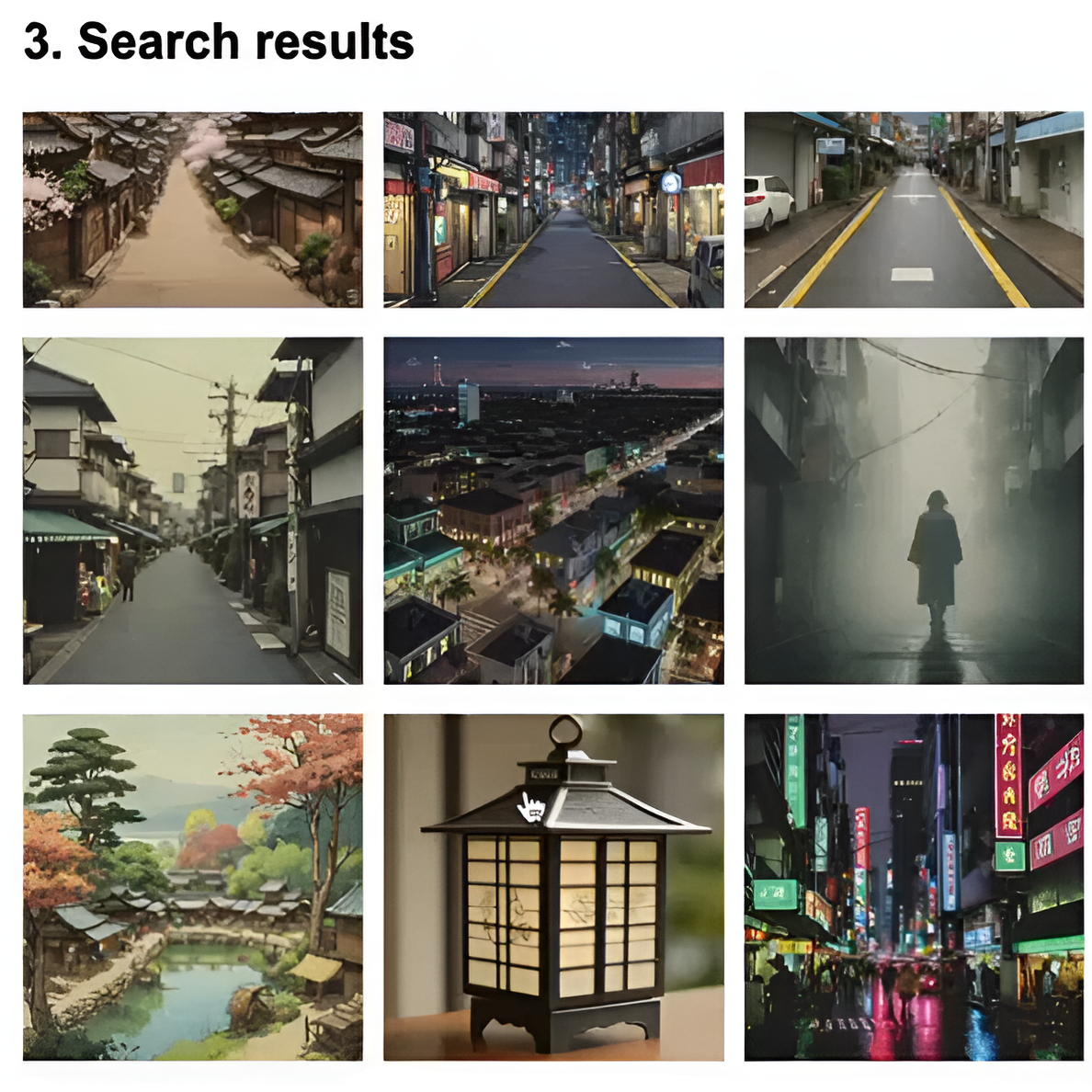}
        \caption{\textsc{Nearest Neighbor}}
        \label{fig:nearest_neighbour}
    \end{subfigure}
    \hfill  
    \begin{subfigure}[t]{0.30\linewidth}
        \includegraphics[width=\linewidth]{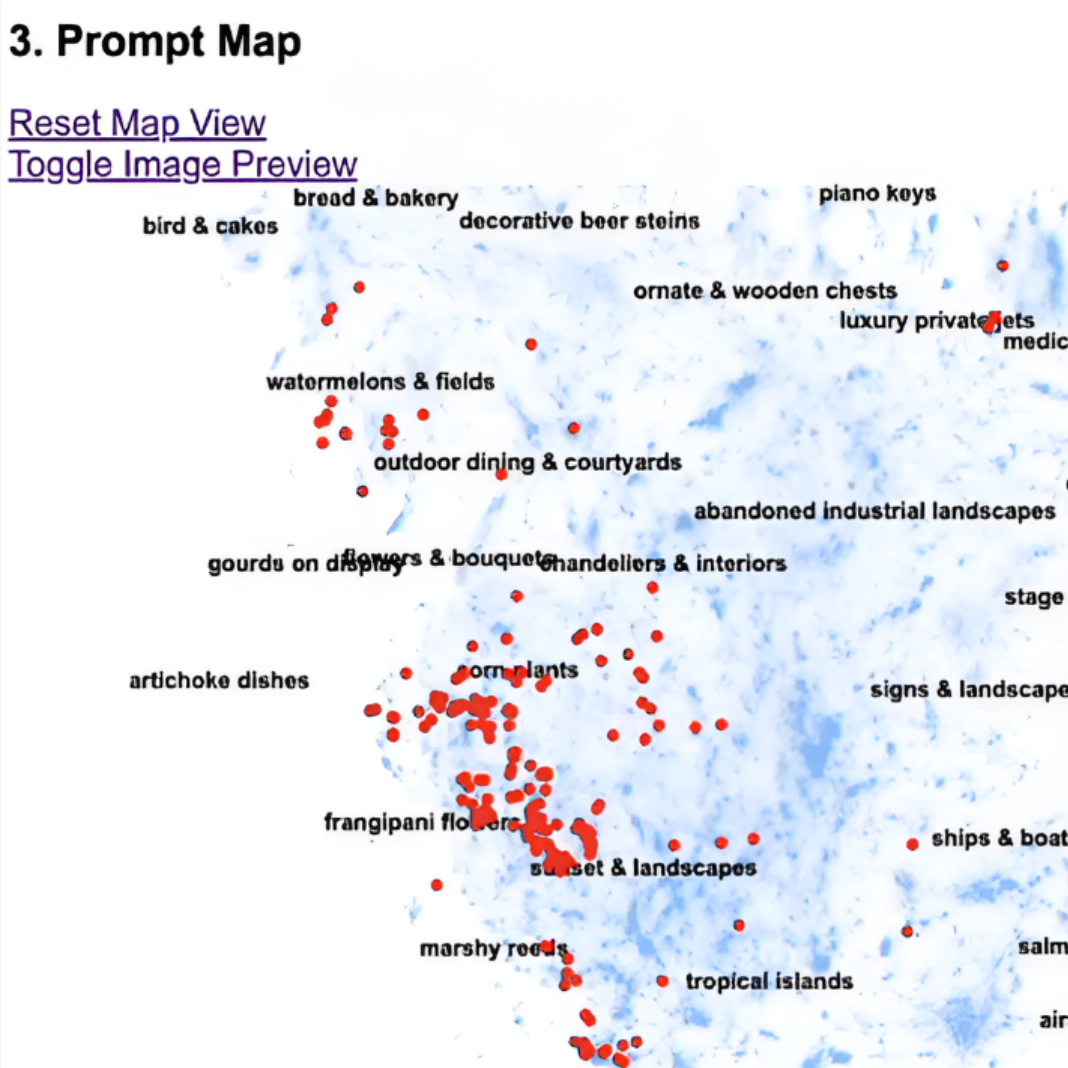}
        \caption{\textsc{PromptMap}}
        \label{fig:prompt_map}
    \end{subfigure}
    \caption{During user studies, participants used the UI shown in \autoref{fig:teaser} and two ablated versions of it. The search bar was removed in the \textsc{No Support} condition, and the tab with generated images replaced the map view. In \textsc{Nearest Neighbor} condition, users could search DiffusionDB~\cite{wang2023diffusiondb} for image examples. The search results were displayed as a grid view that replaced the map. Finally, in \textsc{PromptMap} condition, participants used the full version of the interface and could explore the dataset using the map view and search feature.}
    
    \Description[Screenshots]{Figure compares three different methods for exploring content: No Support, Nearest Neighbor, and PromptMap. Each sub-figure highlights a different approach for selecting or visualizing generated images based on user queries or preferences. Left (a) No Support Shows generated images without additional organizational or exploratory tools. Example images include scenes of interstellar space with gas clouds, galaxies, and stars. Captions provide brief descriptions of the images, such as: "Interstellar space with gas clouds, galaxies, and stars." "Interstellar space with gas clouds, galaxies, and stars with more focus on a planet with a dark shade." Users can delete images or select one as a solution, but there is no further assistance for discovery or refinement. Middle (b) Nearest Neighbor Displays search results in a grid format, showing visually similar or contextually relevant images. The example shows results for a query involving urban and rural landscapes, including Streetscapes with neon lights. Quiet traditional villages. A lone figure walking in a foggy alleyway. A serene garden with autumn colors. This method enables users to find visually similar content but does not map broader thematic relationships. Right (c) PromptMap Visualizes a map of prompts, where red dots represent different prompts with images. Clusters of related topics are labeled, such as: "Outdoor dining and courtyards." "Sunset and landscapes." "Abandoned industrial landscapes." This approach provides a comprehensive and intuitive way to explore diverse content, allowing users to find related topics or navigate between clusters.}
    \label{fig:conditions}
\end{figure*}

\section{Evaluation}

To evaluate how PromptMap contributes to the participant's workflow, we conducted a between the subject quantitative study ($n=60$) and within the subject qualitative study ($n=12$).

\subsection{Quantitative Study}

This section describes the participant population, procedure, and quantitative study results.

\subsubsection{Study Design}

The experiment was structured as a between-subject study. Each participant was assigned to one of three conditions:

\begin{itemize} 
\item \emph{No Support}: Participants used the system version that provided no support in the prompting task. This condition is the default mode of interactions between the text-to-image models and users (see ~\autoref{fig:no_support}). 

\item \emph{Nearest Neighbor}: In this condition, participants had access to a nearest neighbor search feature to find prompts based on a search query. The results were displayed as a grid of images, and users could view the respective prompts by clicking on the images (see ~\autoref{fig:nearest_neighbour}). This type of workflow is present in prompt galleries such as lexica.art~\cite{lexica} or civitai~\cite{civitai}.

\item \emph{PromptMap}: Participants used the entire PromptMap interface, which featured a map-like view for exploring our large-scale collection of prompts. Prompts were displayed in clusters based on similarity, enabling users to explore related examples visually and interactively (see ~\autoref{fig:prompt_map}). 
\end{itemize}

Each participant completed two tasks with no time limit imposed. The task required users to generate an image needed in a given scenario. The order of the tasks was randomized. Two scenarios were used in the study: creating a custom desktop wallpaper and designing an image for a birthday card. Users were not instructed as to the topic of the image and had to arrive at it on their own. Scenarios were chosen during the task elicitation study.

We used the following measures to evaluate performance and user experience across conditions: 
\begin{itemize} 
\item NASA Task Load Index (NASA-TLX)~\cite{hartDevelopmentNASATLXTask1988a}: This measure assessed the perceived workload for each condition, capturing aspects such as mental demand, effort, and frustration. 
\item Creativity Support Index (CSI)~\cite{cherryQuantifyingCreativitySupport2014}: This index measured how well the system supported creativity, focusing on aspects such as exploration, enjoyment, and results worth effort. 
\item UMUX-Lite~\cite{lewis2013umux}: These two questions provide a quick measure of perceived usability and system effectiveness.
\item Custom Satisfaction Questions: Participants were asked to rate the aesthetic quality of their results, how closely the image matched their creative vision, and their overall satisfaction with the generated output. Those questions had the form of a 7-point Likert scale from strongly disagree to agree strongly.
\end{itemize}

Task completion times were also recorded, and each participant's interactions with the system, such as prompt queries or image selections, were logged for further analysis.

\subsubsection{Task Elicitation Study}

To solicit the tasks, we conducted a two-part anonymous task elicitation survey. In part one, we recruited users experienced with generative AI and asked them to provide ideas for practical tasks that can be performed with text-to-image AI models. We distributed the initial elicitation survey among $12$ ML experts and generative AI enthusiasts. All participants reported at least a passing experience with either LLM or image generation. With $8$ participants reporting intermediate or above-average experience with T2I models or LLMs. After removing duplicate ideas, we obtained $19$ ideas for the tasks. Then, an expert familiar with the capabilities of text-to-image models removed the tasks that were not technically feasible, narrowing down the task pool to $7$ tasks.

Next, we conducted a task rating survey using a prolific online research recruitment platform. We presented the tasks as hypothetical scenarios and asked participants to select how likely they would be to use the AI image generator in each scenario. All tasks were rated by $40$ participants drawn randomly from the same participant pool as the main study. In graphics-related activities, $23$ participants reported novice or beginner levels of experience, with $11$ novices and $12$ beginners. Eight participants had intermediate experience, $4$ were advanced, and $5$ described themselves as experts. Experience with text-to-image AI was predominantly novice, with $18$ participants reporting little to no familiarity. Eight participants had beginner-level experience, another $8$ were intermediate, $5$ were advanced, and $1$ participant identified as an expert. Experience with large language models was more evenly distributed. Seventeen participants were novices or beginners, with $10$ novices and $7$ beginners. Eight participants had intermediate experience, another $8$ were advanced, and $7$ described themselves as experts.

After conducting the survey, we selected the two tasks with the highest ratings and used them in our study. The chosen tasks are as follows: 1) Creating a custom desktop wallpaper for your computer based on your favorite book, TV series,  movie, or video game and 2) creating an image for a birthday card for a friend.

\subsubsection{Participants}
We recruited $n=60$ participants through the online research recruitment Prolific. Their age ranged from $19$ to $59$ ($ M = 30.14$, $ SD = 7.46$). The participant genders were distributed as follows: $18$ participants were female, $41$ male, and $1$ non-binary. 

The participants came from diverse fields, with $20$ people working in IT and technology-related roles such as Software Developers, IT Managers, Programmers, and IT Consultants. Five participants were from engineering fields, and another five worked in education as teachers or tutors. Three participants were unemployed or homemakers, while three worked in retail or store management. Additionally, $3$ came from creative fields like graphic design, digital arts, and performing arts. There were also participants from business and marketing ($3$), public service ($2$), and service-oriented jobs like catering, janitorial, and warehouse work ($3$). 

The participants' visual art experience was varied, with $37$ participants reporting having no experience or beginner level of experience and $18$ participants reporting intermediate experience. The population also contained $4$ participants with advanced experience and a single expert. Regarding the experience with the image-to-text generation, $34$ participants reported having beginner experience or no experience with the image-to-text generation, $21$ reported intermediate experience, $5$ reported advanced experience, and $1$ participants reported being an expert. Experience with Large Language Models was reported as beginner level by $15$ participants, intermediate level by $18$ participants, advanced level by $16$ participants, and expert level by $5$ participants. Participants reported low experiences with text-to-image generation at $M=1.35$ ($SD=0.97$)\footnote{On a 5-point Likert scale from "No experience" to "Expert".}. We found no significant differences in experience between the three conditions.

Each participant received compensation for their involvement at a rate of £$8.25$/hr --- a recommended rate by Prolific, with a median study completion time of $32min$. Before entering the study, participants were informed that they were required to be adults, fluent in English, and heaving normal or corrected to normal vision.

To ensure the quality and relevance of the data gathered from Prolific, we apply attention checks in the form of additional questions where there is only one objectively correct answer or response, e.g., "for this question, select \textit{Strongly agree}". This allows us to confirm that participants carefully read the questions Following Prolific's policy~\footnote{https://www.prolific.com/resources/weve-updated-our-attention-and-comprehension-check-guidance}, each participant was presented with five attention checks, and failing two resulted in rejection. No participants were rejected based on the attention check failure.

\subsubsection{Procedure}
The study was conducted in a between-subject setup. The $60$ participants were randomly split into one of three conditions: 1) \emph{No Support}, 2) \emph{Nearest Neighbor}, and 3) \emph{PromptMap}. Each condition was performed by $20$ participants.

Once the participants accepted the study on the prolific platform, they were redirected to the external website containing the consent form, data protection information, and the demographic survey. Upon completing the demographic survey, they were redirected to the prototype of the system to start the first phase of the experiment: a brief text-to-image tutorial. 

The participants were instructed first to create a short caption describing a scene and then, in each step of the tutorial, to modify the different aspects of the image (ex., medium or lighting). After each modification, they were instructed to generate an image. Upon completing the text-to-image warm-up, participants were instructed to watch a short video tutorial regarding the user interface assigned to them and the task. 

The participants were then redirected to the user interface. Each participant solved two tasks using the interface assigned to their condition. The order of tasks was randomized. Beside the scenario from the task elicitation study, no suggestions as to the possible image topics were given to the participants. Open-ended tasks were used to improve the ecological validity of the study. To enforce a minimum level of engagement with the user interface, the ability to select the image as a submission was unlocked after three different prompts were used to create images. We did not enforce any limit on time or the number of generated images. We found that $37$ out of $60$ users made more than three unique prompts per task. The video tutorial instructed participants to select the image as a solution and submit it using the submission button once they arrived at a satisfactory result.

After completing both tasks, participants were redirected to the external survey website. They were presented with NASA-TLX~\cite{hartDevelopmentNASATLXTask1988a}, Creativity Support Index (CSI)~\cite{cherryQuantifyingCreativitySupport2014}, and UMUX-lite~\cite{lewis2013umux} questionnaires.  We additionally added three custom questions to gauge participants' satisfaction with the results. We asked 1) whether the result was aesthetically pleasing, 2) whether the result matched their artistic vision, and 3) whether they were satisfied with the result. We also included a text field for any additional comments. Upon completing the survey, participants were redirected back to the prolific platform to complete the task and claim the remuneration.

\subsubsection{Results}
For all quantitative results, we conducted one-way ANOVAs after rank-aligning the data (if normality was violated). ART-ANOVA is an established,  robust alternative to standard non-parametric approaches ~\cite{wobbrockAlignedRankTransform2011}.

\begin{figure}[h]
    \centering
    \includegraphics[width=\columnwidth]{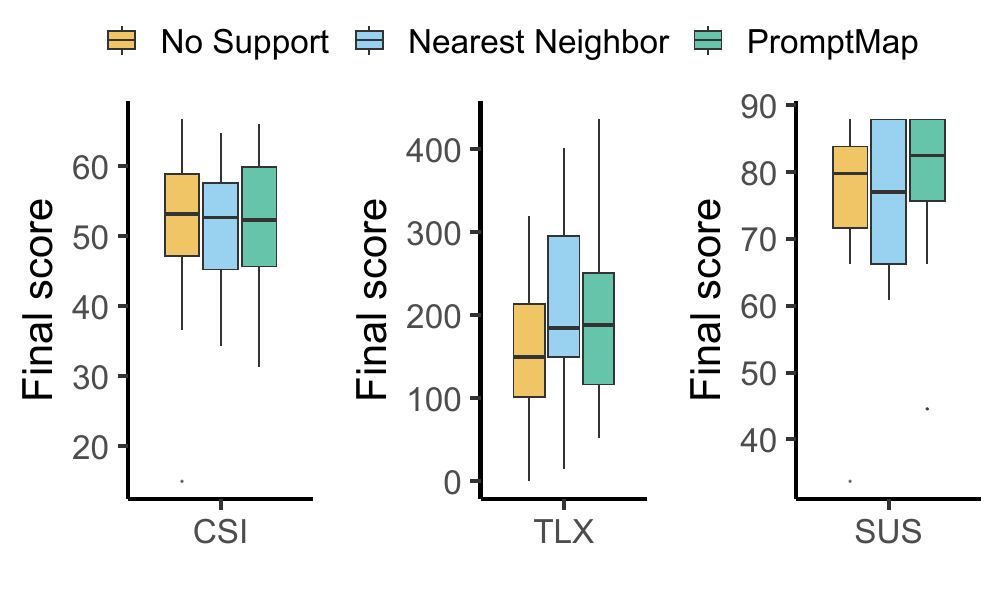}
    \caption{Final scores as calculated based on CSI~\cite{cherryQuantifyingCreativitySupport2014}, NASA-TLX~\cite{hartDevelopmentNASATLXTask1988a} and UMUX-Lite~\cite{lewis2013umux}. The latter has been provided as SUS~\cite{lewis2015investigating} parity score. No significant differences between the conditions were found.}
    \Description[Boxplot]{Figure presents box plots to compare the performance of three methods—No Support, Nearest Neighbor, and PromptMap—across three evaluation metrics: CSI, TLX, and SUS.}
    \label{fig:questionnaires}
\end{figure}

\paragraph{CSI, NASA-TLX, UMUX-Lite}
For neither the CSI~\cite{cherryQuantifyingCreativitySupport2014}, the NASA-TLX~\cite{hartDevelopmentNASATLXTask1988a}, nor the UMUX-Lite~\cite{lewis2013umux} scale did we find any significant difference between our conditions. Individual sub-scales for CSI and NASA-TLX also did not show any significant difference. \autoref{fig:questionnaires} provides a visual representation of the three scales' results. We display the UMUX-Lite as the SUS parity score for comparison~\cite{lewis2015investigating}. According to the rating scale by \citet{bangor2009determining}, this places all systems in the "good" category for system usability with \emph{PromptMap} slightly in the lead. All systems score equally on the creative support questionnaire, yet \emph{Nearest Neighbor} and \emph{PromptMap} score higher on NASA-TLX, thus requiring more effort from the user to complete the task.

\begin{figure}[h]
    \centering
    \includegraphics[width=\columnwidth]{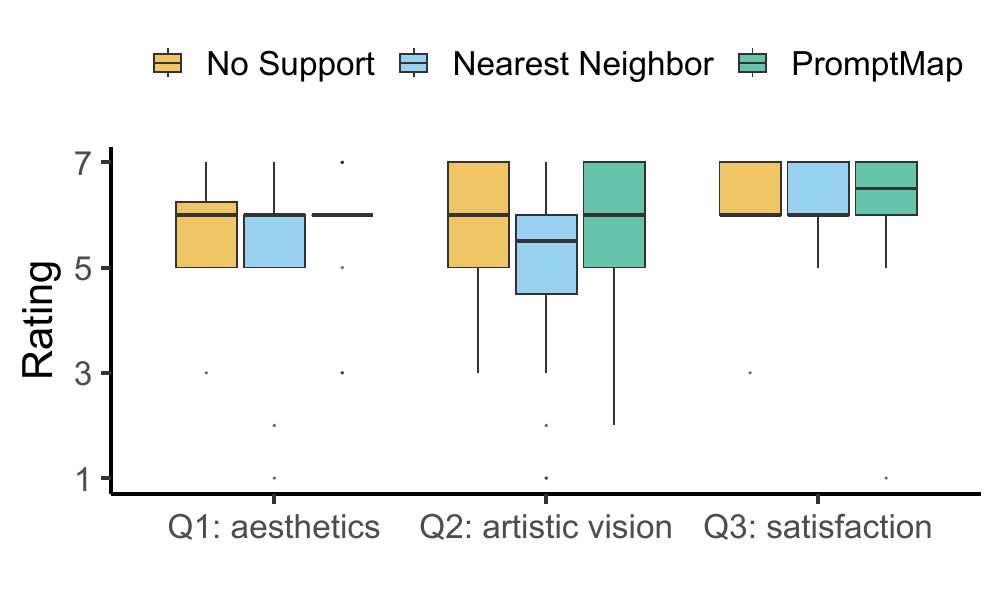}
    \caption{Ratings for our custom questions inquiring about aesthetics (Q1), matched artistic vision (Q2), and satisfaction with the final image results (Q3) given our conditions.}
    \Description[Boxplot]{Figure presents box plots to compare the performance of three methods—No Support, Nearest Neighbor, and PromptMap—across three subjective evaluation questions: aesthetics, artistic vision, and satisfaction.}
    \label{fig:custom_questions}
\end{figure}

\paragraph{Custom Questions}
We found no significant effects of the image generation methods on our custom questions. The results are depicted in \autoref{fig:custom_questions}.

\begin{figure}[h]
    \centering
    \includegraphics[width=\columnwidth]{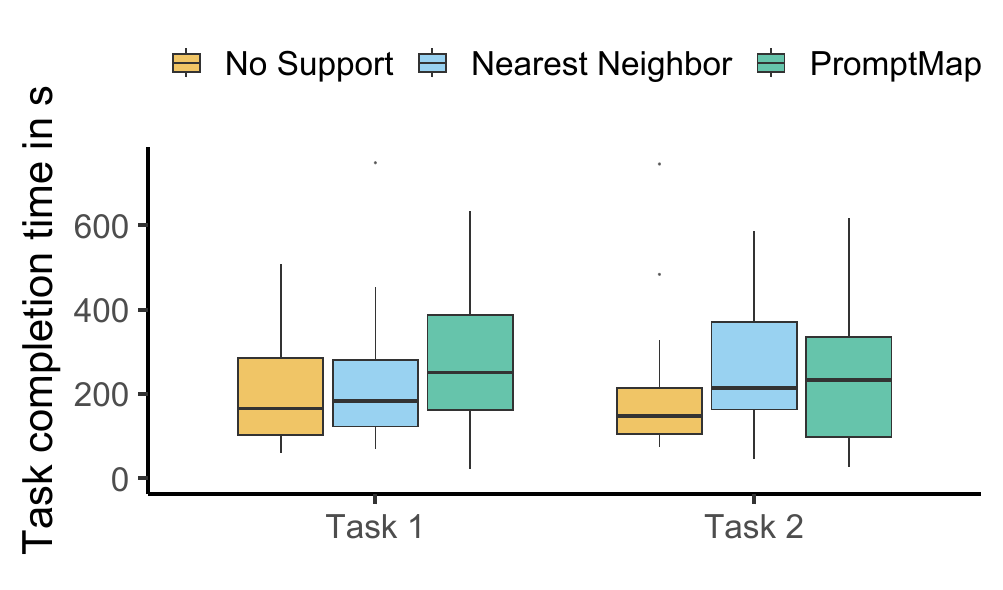}
    \caption{Task completion times for the two tasks given our conditions.}
    \Description[]{Figure Present box plots to compare the task completion time (in seconds) for three methods—No Support, Nearest Neighbor, and PromptMap—across two tasks: Task 1 and Task 2.}
    \label{fig:tct}
\end{figure}

\paragraph{Task Completion Times}
We analyzed the task completion times for the participants for both tasks. The results are visualized in \autoref{fig:tct}. We found no significant differences between the two tasks, given the conditions. On average (combining both tasks), participants were faster for \emph{No Support} ($M=408\,s$, $SD=216\,s$), followed by \emph{Nearest Neighbor} ($M=549\,s$, $SD=295\,s$) and took longest for \emph{PromptMap} ($M=623\,s$, $SD=378\,s$).

\subsection{Qualitative Study}

This section describes the participant population, procedure, and qualitative study results.

\subsubsection{Study Design}

We conducted a qualitative study to gain insight into the effect of PromptMap on user's workflow. The experiment was structured as a within-subject study to allow participants to compare different interface versions. The study evaluated the same conditions as the quantitative study: \emph{No Support}, \emph{Nearest Neighbor}, and \emph{PromptMap}. The same prototypes were used. To accommodate time constraints, we modified the experiment procedure from the quantitative study. We replaced the text-to-image tutorial with a short introduction with examples shown by the researcher. We also reduced the number of tasks to a single task: wallpaper creation. Participants were instructed to produce three different wallpapers on three different topics.

Additionally, we introduced a 10-minute time limit for each interface. The order in which participants used the UIs was randomized using Latin Square. At the end of the study, we conducted semi-structured interviews during which we asked questions about 1) What was their process of arriving at the results, 2) what were the main challenges they encountered, 3) whether they had a clear vision of what they wanted to create at the beginning of the process, 4) the advantages and disadvantages of the shown interfaces, 5) how they used examples in their creative process, 6) as to which example set they preferred and why.

\subsubsection{Participants}
We recruited participants using university mailing lists, social media, and snowball sampling. In total, we recruited $n=12$ participants ($4$ female, $7$ male, $1$ non-binary) aged between $22$ and $39$ years old ($M = 28.27, SD = 4.30$). 

The participants primarily came from academic and research-oriented fields, with $7$ being engaged in research and academia, including computer science researchers, AI researchers, HCI researchers, and lecturers/teachers. The remaining $5$ participants were involved in engineering or programming.

The participants represented a diverse range of visual art experience, with $5$ reporting intermediate experience, $4$ identifying as beginners, $1$ reporting advanced expertise, and $2$ reporting no experience in the field of graphic design or visual art.

For image-to-text generators, $5$ participants reported beginner experience, $5$ reported intermediate experience, $1$ reported advanced experience, and $1$ participants reported no experience with tools like Stable Diffusion or DALL·E.

Regarding Large Language Models, $5$ participants reported an intermediate level of experience, $3$ reported advanced experience, $1$ reported being an expert, and $3$ participants identified as beginners.

\subsubsection{Procedure}

Participants were invited to join an online call with the researcher. First, they were asked to fill out a demographic survey. Then, the researcher gave a short presentation containing task instructions and examples of prompts. During the study, participants were tasked with creating three different desktop wallpapers, one per condition. Participants were explicitly asked to use different topics in each condition. Participants had up to 10 minutes with each interface version and were allowed to ask questions pertaining to the user interfaces during the study. The researcher introduced each interface with a short description of each component. At the end of the experiment, a brief semi-structured interview was conducted with each participant. We recorded (total duration 3~h 17~m) the interviews and transcribed them verbatim. We used the pragmatic approach to the thematic analysis described by Blandford et al.~\cite{blandford2016qualitative}. An inductive coding strategy was used to obtain the initial set of codes. Two researchers independently coded four representative samples (30\% of the material) to establish the initial codebook. We then organized a collaborative session where we derived a coding tree based on the open-coded dataset. Two researchers then re-coded the entire data corpus.

\subsubsection{Results}

We present our findings organized into two themes: a) \textsc{Creation Process}, addressing the exploration methods, sources of inspiration, possible system support, as well as challenges; and b) \textsc{The role of examples}, describing user's preferences and outcomes related to supporting prompt creation by visual and prompts examples. We support our findings with participants' quotes, \textit{highlighted in italics} and identified by participant IDs along with the respective condition they refer to.

\textbf{Creation Process} During the process of creating prompts, most participants identified two strategies in their approach: standard iterative writing associated with the \textsc{No Support} condition and example-driven approach in the \textsc{Nearest Neighbor} and \textsc{PromptMap} condition.

In the case of the condition without any support, participants often generated images through trial and error, frequently regenerating the same prompt to gather insights on how to adapt it to achieve the desired results.

\begin{quoting}
(P12\textsubscript{No Support}) ...to build this image, I had to add things by trial and error
\end{quoting}

\begin{quoting}
(P3\textsubscript{No Support}) I'm writing a prompt to see how it changes with adding and getting off some random words ... just add one word, see what it generates, another one, and see what it generates.
\end{quoting}

When introduced to approaches with examples---\textsc{Nearest Neighbor} and \textsc{PromptMap}, participants shifted their focus from relying on the iterative refinement of prompts through small changes to the exploration of visual examples. They favored exploring the image and prompt examples and using them as a source of inspiration and guidance before analyzing and refining their prompts.

\begin{quoting}
(P1\textsubscript{Nearest Neighbor}) I gave an initial prompt of what I wanted, more or less, and then scanned through the options, chose the one that came closest, and then I went back to modifying the prompt to get me closer to where I want to be.
\end{quoting}

\begin{quoting}
(P3\textsubscript{PromptMap}) ...to just be able to go over it and see random stuff, some random ideas, like those general prompts or things that I like...
\end{quoting}

\begin{quoting}
(P4\textsubscript{Nearest Neighbor}) So the ones that I liked and then I checked their prompt and got some ideas from it and we reused them in my own prompt.
\end{quoting}

In the example-driven approach, participants often began with short, general search queries and then explored the search results to find a direction aligned with their intended result. As they refined the prompts, they gained a clearer understanding of how specific elements influenced the outcomes. This process helped them better grasp the link between the prompt and the resulting image output, making it easier to produce the desired result.

\begin{quoting}
(P1\textsubscript{Nearest Neighbor}) I gave an initial prompt of what I wanted, more or less, and then scanned through the options, chose the one that came closest, and then I went back to modifying the prompt to get me closer to what I want.
\end{quoting}

When using the \textsc{PromptMap}, participants appreciated that the prompt examples reduced the need to find precise wording for getting the result matching their expectations. Furthermore, participants emphasized the importance of inspiration and valued PromptMap for the ability to explore images with similar themes, which supported the process of writing and refining prompts. They particularly appreciated how the visual grouping of various topics introduced ideas beyond their initial scope, facilitating the broader exploration of ideas.

\begin{quoting}
(P1\textsubscript{PromptMap}) I mostly just worked with the map. I think I typed one thing into the search to get some of the point clouds highlighted to better see where I want to go or where the relevant stuff for me is. But then I mostly use the point cloud and to find a prompt, use this prompt, and then just slightly modified it.
\end{quoting}

\begin{quoting}
(P6\textsubscript{PromptMap}) I want to explore what there is to find and what groups are similar and where can I find the images.
\end{quoting}

\begin{quoting}
(P1\textsubscript{PromptMap}) I think I played around with the prompt the most, and the more different options I got, the less I was forced to just change the wording in the prompt slightly because I was now able to explore what possibilities there are and how I wanted to use them. 
\end{quoting}

\textbf{The role of examples} Participants frequently pointed out the challenges of creating text-based prompts for current image generators. They emphasized the difficulty of predicting the outcome due to a lack of understanding of the generation mechanisms and how changes in the prompt would ultimately affect the resulting image. Some also noted the impossibility of achieving a perfectly tailored result, leading to a sense of losing control over the outcome and the inability to edit smaller parts of the image, frustration with generated errors, and the difficulty of crafting an ideal prompt.

\begin{quoting}
(P7\textsubscript{No Support}) I had rather problems with some areas of the image which weren't quite right. So the image was overall good. But then there's a small error where you can directly see...
\end{quoting}

\begin{quoting}
(P10\textsubscript{No Support}) ...I don't feel much control over what I'm producing ... maybe I'd have to depreciate prompts even more
\end{quoting}

\begin{quoting}
(P9\textsubscript{No Support}) The main challenge was getting system to not ignore certain parts of the prompt.
\end{quoting}

In the case of \textsc{Nearest Neighbor} and \textsc{PromptMap}, participants noted that examples were helpful in the visual exploration phase, where they began with general prompts and then scanned the generated images to identify a direction that aligned with their creative intentions. Rather than starting from scratch, they could draw inspiration from pre-existing images or prompts and modify them to suit their needs. They mentioned that this approach allowed them to refine their understanding of how different elements impacted the image, ultimately leading to better results. Participants noted that some prompts suggested descriptions containing elements they hadn't previously considered possible. Hence, they suggested that an assistant, functioning like a copilot or a questionnaire guiding them through elements of scene, style, and color, would be highly beneficial for prompt creation.

\begin{quoting}
(P12\textsubscript{PromptMap}) I really like how prompts are described, because it reminds me that I can type in cool colors, sci-fi genre ... things that I completely forgot to type... it gives me more, system reminds me, "Hey, you can actually add different things and they change".
\end{quoting}

The \textsc{Nearest Neighbor} and \textsc{PromptMap} were particularly appreciated for their ability to present examples in a visually organized manner. Participants reported that the visual clustering of related themes and topics made exploring ideas they had not initially considered easier.

\begin{quoting}
(P4\textsubscript{PromptMap}) And then these red dots to see how they were classified to specific category made it easy to decide whether I wanted to explore it to see what they generate in that area.
\end{quoting}

Finally, participants remarked that the synthetically generated prompts shown in the \textsc{PromptMap} serve as a more detailed description of the example images compared to the prompts in the \textsc{Nearest Neighbor} condition. Additionally, participants remarked that examples shown in \textsc{Nearest Neighbor} condition demonstrated less variety, often perceiving it as monotonous, more general, and less helpful for broad alternative exploration compared to \textsc{PromptMap}.

\begin{quoting}
(P7\textsubscript{Nearest Neighbor}) I see that the prompts are not very long. There are short prompts. I don't know which captioning tool is used, but probably there could be a bit more details, but otherwise mind, there's a list I can scroll and that's probably good. I don't see anything else you could add or remove. 
\end{quoting}

\begin{quoting}
(P4\textsubscript{PromptMap}) I think in terms of variety the database that fed the map could have been more helpful...
\end{quoting}

Higher diversity of synthetic prompts and preference for them was also indicated by participants.
\begin{quoting}
(P1\textsubscript{PromptMap}) You had a lot of just library with people, with robots, without people, with animals, with whales, with whatever. And I was just able to scan through this and see a lot of just different stuff. That might be nice.
\end{quoting}

\begin{quoting}
(P7\textsubscript{PromptMap}) [Why did you prefer the synthetic prompts?] Mostly because the prompts were longer and had more details in it.
\end{quoting}

\section{Discussion}
Our study demonstrates an alternative, example-driven workflow for creating text-to-image images. In related interfaces~\cite{brade2023promptify,wang2024promptcharm,Feng2024promptmagician}, the weight has been placed on the recommendation of keywords~\cite{Feng2024promptmagician}, prompts~\cite{brade2023promptify} and improving explanability using attention maps~\cite{wang2024promptcharm}. We show that providing users with varied examples positively impacts their process and helps them avoid arriving at their solution through trial-end-error changes to the prompt. 

PromptMap is designed to support users during interactions with text-to-image models by allowing users to explore a large dataset to find relevant examples. While the results of our qualitative study show that many participants regarded examples as useful, we did not observe significant differences in the measured parameters between PromptMap and our two baselines. CSI~\cite{cherryQuantifyingCreativitySupport2014}, TLX~\cite{hartDevelopmentNASATLXTask1988a} and UMUX-lite~\cite{lewis2013umux} scales did not record statistically significant differences. All conditions are placed in the "good" category for system usability with \emph{PromptMap} slightly in the lead.

\subsection{Presence of Examples Changes User Strategies}
The results of the qualitative study, on the other hand, revealed that the presence of examples in \emph{Nearest Neighbor} and \emph{PromptMap} conditions changed the creative process employed by the participants. We identified two strategies applied by the participants depending on whether the examples were present or not.

In the \emph{No Support} condition, participants relied heavily on a trial-and-error approach, where they made small, incremental changes to prompts to observe how they influenced image outputs. This approach often led to frustration, as participants struggled to control specific aspects of the images and experienced difficulties refining their prompts to achieve desired outcomes.

In contrast, introducing examples in the \emph{Nearest Neighbor} and \emph{PromptMap} conditions changed participants' workflow from trial-and-error to a more example-driven approach. Participants often searched for examples related to their topic to identify the best-fitting sample and then used it as a starting point for their prompt. We also found that scanning through visual and textual examples gave participants useful examples of how prompts can be structured.

Furthermore, we found that the organization of examples is vital for supporting the exploration process. Participants appreciated the visual clustering and thematic organization in PromptMap and reported that it facilitated their discovery of new ideas beyond their initial scope.

\subsection{Generating Synthetic Prompt Datasets is Feasible}
Our synthetic prompt generation pipeline showed potential as a method for generating large collections of varied prompt examples without the need for data scraping. Our results indicate that although our synthetic prompts lack modifiers and are more concise, they are equally or even more preferred by users than the human prompts collected in the wild. 

Curiously, we found that even though prompts in our synthetic dataset lack "magic keywords" ~\cite{oppenlaender2023taxonomy, dehouche2023s} and are on average shorter than prompts in DiffusionDB~\cite{wang2023diffusiondb}, they were perceived as being more detailed and varied. This observation, coupled with a lack of significant change in reported satisfaction from the output between \emph{Nearest Neighbor} and \emph{PromptMap} conditions, suggests that more concise image captions perform the same or better as long examples with numerous keywords describing the style of the image (modifiers). 

\subsection{Limitations and Future Work}

With PromptMap, we proposed a new interaction style for AI-based image generation. Since our work aims at circumventing elaborate prompt engineering by design, it does not teach users to learn the intricacies of image prompts. In the future, we envision combining both methods, additionally integrating other support tools such as attention map visualization~\cite{wang2024promptcharm} and other control inputs such as segmentation masks~\cite{couairon2023zero} or inpainting~\cite{xie2023smartbrush}. Additionally, both the data generation pipeline and our interaction method can be applied to other types of text-guided generation, such as text-to-3D, text-to-video, or text-to-audio. The application of our method to those domains can be investigated in the future.

In our dataset, we made a deliberate choice not to include specialized keywords (ex. '4k', 'artstation', 'greg rutkowski') in prompts as reliance on them can lead to prioritization of surface-level aesthetical qualities and reduce the overall diversity and innovation \cite{palmini2024patternscreativityuserinput}. Future work could investigate how to recommend keywords to address this challenge.

During the interviews, participants reported issues with the map's usability. Difficulty in selecting points, displaying too many results, and having to select points to see the images were frequently brought up by participants. Participants suggested that the display similar to the grid shown in \emph{Nearest Neighbour} condition combined with thematic organization and example variety of \emph{PromptMap} would improve usability significantly. Better label placement and image preview selection algorithms can be investigated in future work. Additionally, the output from UMAP~\cite{2018arXivUMAP} contains both regions of significant point density and empty space. We also acknowledge that although we did not encounter significant complaints about topic placement, the alignment between the distances in the "topic space" shown on the map and the human perception of topic similarity is unknown. The creation of algorithms for laying out the points and selecting representative samples according to human perception are promising future avenues for research.

We also encountered several limitations in the prompt generation pipeline. Although not remarked on by the participants, during the inspection of the dataset, we discovered that some prompts describe very niche or unusual combinations of topics that the text-to-image models do not correctly generate. An additional filtering step can be considered to detect and remove such examples from the dataset. We also discovered that some prompt annotations do not match the image output image due to them being predictions made from the prompt and not the output image from the text-to-image model. This problem can be mitigated by choosing a more capable image generator or using VLLM to annotate the images. Finally, due to resource constraints, the detailed investigation of the influence of generation parameters on the pipeline behavior remains a topic for future research.

\section{Conclusions}

In this paper, we report on the implementation and evaluation of PromptMap, a user interface that allows users to freely explore an extensive, synthetic collection of prompt examples as part of their workflow. We evaluated the interface on crowd-sourced image creation tasks. We find that the ability to freely explore a vast collection of examples shifts users away from trial-and-error prompt creation and towards example-driven approaches. Our qualitative results highlight that varied and thematically organized examples shown in our PromptMap interface positively contribute to their workflow. We also demonstrate the feasibility of fully synthetic large prompt collections. This work contributes to the understanding of how the exploration of examples can enhance methods for interaction with text-to-image models.

\bibliographystyle{ACM-Reference-Format}
\bibliography{main}

\appendix

\end{document}